\documentclass[amsmath,amssymb,article,preprints,article,accept,moreauthors,pdftex,10pt,a4paper]{revtex4-1}

\usepackage[utf8x]{inputenc}
\usepackage{ucs}

\usepackage{booktabs}

\usepackage{verbatim}
\usepackage{amsfonts}
\usepackage{dcolumn}% Align table columns on decimal point
\usepackage{bm}% bold math
\usepackage[usenames]{color}
\usepackage{multirow}
\usepackage{graphicx}
\usepackage{hyperref}
\usepackage{subfig}
\usepackage{booktabs}
\usepackage{xcolor}
\usepackage[normalem]{ulem}

\usepackage{ucs}
\usepackage{amssymb}
\usepackage{makeidx}
\usepackage{diagbox}
\usepackage{tabularx}
\usepackage{amsthm}

\usepackage{float} % Allows putting an [H] in \begin{figure} to specify the exact location of the figure
\usepackage{wrapfig} % Allows in-line images such as the example fish picture

\usepackage{upgreek} % para poner letras griegas sin cursiva.
\usepackage{cancel} % para tachar.
\usepackage{mathdots} % para el comando \iddots
\usepackage{mathrsfs} % para formato de letra
\graphicspath{{figures/}} % Specifies the directory where pictures are stored

\begin{document}

%\preprint{draft version}

\title{Entropic analysis of the quantum oscillator with a
	minimal length}

\author{D. Puertas-Centeno$^{1}$ and
%\footnote{Corresponding Author: arplastino@ugr.es}
M. Portesi $^{2,3}$}
%\shortauthor{I.V. Toranzo \etal}

\address{
  $^{1}$ \quad  Departamento de Matem\'atica Aplicada, Universidad Rey Juan Carlos, 28933 Madrid, Spain; david.puertas@urjc.es\\
  $^{2}$ \quad Instituto de F\'isica La Plata (IFLP), CONICET, 1900 La Plata, Argentina\\
  $^{3}$ \quad Facultad de Ciencias Exactas, Universidad Nacional de La Plata (UNLP), 1900 La Plata, Argentina; portesi@fisica.unlp.edu.ar}

\begin{abstract}
The well-known Heisenberg--Robertson uncertainty relation for a pair of noncommuting observables, is expressed in terms of the product of variances and the commutator among the operators, computed for the quantum state of a system. Different modified commutation relations have been considered in the last years with the purpose of taking into account the effect of quantum gravity. 
Indeed it can be seen that letting $[X,P] = i\hbar (1+\beta P^2)$ implies the existence of a minimal length proportional to $\sqrt\beta$. 
The Bialynicki-Birula--Mycielski entropic uncertainty relation in terms of Shannon entropies is also seen to be deformed in the presence of a minimal length, corresponding to a strictly positive deformation parameter $\beta$. 
Generalized entropies can be implemented. Indeed, results for the sum of position and (auxiliary) momentum R\'enyi entropies with conjugated indices have been provided recently for the ground and first excited state. We present numerical findings for conjugated pairs of entropic indices, for the lowest lying levels of the deformed harmonic oscillator system in 1D, taking into account the position distribution for the wavefunction and the actual momentum.
\end{abstract} %end of abstract
\maketitle

\section{Introduction}

One of the pillars in the building of quantum physics is the uncertainty principle, which has been  formulated by Heisenberg \cite{Hei1927} and originally given in terms of the product of variances of position and momentum observables as quantifiers of the quantum particles' spreading. Recently, nontrivial relations have been obtained for the sum of variances~\cite{Mac14}. Besides, it has been proven that  uncertainty relations  can be formulated as well in terms of Shannon and R\'enyi information entropies (see, for instance, \cite{Bia75,Maa88,Zozor2008,Bia11} and references therein). The possible influence of gravity in  uncertainty relations has been recently proposed \cite{Kem95,Pedram2012a,Pedram2012,Hossenfelder2013}, and a modification to Heisenberg  inequality known as generalized uncertainty principle (GUP) has been analyzed.  The modification of the position--momentum uncertainty relation, which is carried out through a deformation of the typical commutation relation between operators, is linked to the existence of a minimal observable  length. It is interesting to quote that an experimental procedure to detect these possible modifications has been proposed~\cite{Pik12}, however it is not yet possible to achieve the required precision.

We consider here the harmonic oscillator and study the deformed uncertainty relations appealing to R\'enyi entropies. In Refs.~\cite{Chang2002, Ped16}, the wavefunctions for some values of the principal quantum number have been given in momentum space, and in position space. However, differently to the standard quantum mechanics, the presence of gravity induces that position and momentum space wavefunctions are not related via a Fourier transform, but it is necessary to consider an auxiliary transformation~{\cite{Ras17}. 
	We present an entropic analysis in this case, and show a family of inequalities satisfied by R\'enyi entropies.

	%%%%%%%%%%%%%%%%%%%%%%%%%%%%%%%%%%%%%%%%%%
	\section{Results}

	%%%%%%%%%%%%%%%%%%%%%%%%%%%%%%%%%%%%%%%%%%
	\subsection{Quantum oscillator wavefunctions with a minimal length} 
	\label{ssec_wavefunction}
	
	The harmonic oscillator (HO) is one of the most relevant physical systems, and it is one of the few quantum systems for which the spectrum and corresponding wavefunctions are exactly known. In general, the Hamiltonian of the 1D HO is given by $H=P^2/(2m)+(m\omega^2/2) X^2$, where $X$ and $P$ are such that their commutator gives the $c$-number $i\hbar$. 
	
	In the context of GUP, a deformation of the standard commutation relation is assumed. 
	Here we assume the form \  $[x,k]=i\hbar(1+\beta k^2)$, with $\beta$ being a positive parameter, which imposes a minimal value for the variance in position, $\Delta X_{\min}=\hbar \sqrt\beta$, of the order of Planck's length. We mention that more general deformations (in arbitrary dimensions) have been proposed~\cite{Kem95}, however this will be focus of further study to be considered elsewhere. 
	
	Under these assumptions, the Schr\"odinger equation in the (auxiliary) momentum space has been solved by Pedram~\cite{Ped16}. Letting $k=\frac{\tan(\sqrt\beta \, q)}{\sqrt\beta}$, the wavefunctions are given by 
	\begin{equation}
	\label{MWF}
	\phi_n(q)=N_n\,C_n^{(\lambda)}\left(\sin\left(\sqrt{\beta}\,q\right)\right) \ \cos^\lambda\left(\sqrt\beta \,q\right) , 
	\qquad q\in\left(-\frac{\pi}{2\sqrt \beta},\frac{\pi}{2\sqrt \beta}\right), 
	\end{equation}
	where $N_n$ is the normalization constant, given by $N_n=\left(\frac{\sqrt\beta \, \Gamma(\lambda)^2\Gamma(n+1)(n+\lambda)}{\pi2^{1-2\lambda}\Gamma(n+2\lambda)}\right)^\frac12$ \ for $n=0,1,2,\ldots$; the symbol $C_n^{(\lambda)}(\cdot)$ denotes the Gegenbauer polynomials, 
	with $\lambda=\frac12\left(1+\sqrt{1+\frac{4}{\eta^2}}\right)$ \ and $\eta=m\hbar\omega\beta$; and the energy levels are given by $E_n=\hbar\omega\left(n+\frac12\right)\left(\sqrt{1+\frac{\eta^2}{4}}+\frac\eta2\right)+\frac12\hbar\omega n^2\eta$. table
	Note that in the limit $\beta\to0^+$, or equivalently $\lambda\to+\infty$, the standard case is recovered as it is shown in \cite{Ped16}. From Eq.~\eqref{MWF} one can compute the position wavefunction through the Fourier transform as 
	\begin{equation}\label{Fourier}
	\psi_n(x)=\frac1{\sqrt{2\pi}}\int_{-\frac{\pi}{2\sqrt\beta}}^{\frac{\pi}{2\sqrt\beta}}e^{iqx}\phi_n(q)\,dq .	
	\end{equation}
	In Ref.~\cite{Ped16} this has been computed exactly for $n=0$ and $n=1$.
	
	As claimed by Rastegin~\cite{Ras17}, the physically legitimate wavefunction in momentum space, which must depend on $k$, is not given by Eq.~\eqref{MWF}, but can be obtained through the following condition: 
	\begin{equation}
	|\tilde \phi_n(k)|^2\,dk=|\phi_n(q)|^2\,dq . 
	\end{equation} 
	Therefore 
	\begin{equation}\label{transformation}
	|\tilde \phi_n(k)|^2=\frac{|\phi_n(q)|^2}{1+\beta\,k^2},
	\end{equation} 
	with \ $q(k)=\frac{\arctan(\sqrt\beta\,k)}{\sqrt\beta}$. Finally 
	\begin{equation}
	\tilde \phi_n(k)=N_n\, C_{n}^{(\lambda)} \left(\frac{\sqrt\beta\,k}{\sqrt{1+\beta\,k^2}}\right)\,(1+\beta k^2)^{-\frac{\lambda+1}2} .
	\end{equation}

	\subsection{Behavior of the sum of R\'enyi entropies}
	\label{ssec_Renyi}
	
	Shannon entropy in momentum space was analytically calculated by Pedram \cite{Ped16} for the ground state and for the first excited state. Here the R\'enyi entropy $R_{\alpha}$ using both representations in momentum space, is numerically studied  for the ground state and the first five  excited ones. 
	R\'enyi entropies are given by 
	\begin{equation}\label{Renyigamma}
	R_\alpha[\phi_n]=\frac1{1-\alpha}\ln\int_{-\frac{\pi}{2\sqrt{\beta}}}^{\frac{\pi}{2\sqrt{\beta}}} |\phi_n(q)|^{2\alpha}\,dq,\quad\text{and}\quad R_\alpha[\tilde\phi_n]=\frac1{1-\alpha}\ln\int_{-\infty}^\infty |\tilde\phi_n(k)|^{2\alpha}\,dk, 
	\end{equation}
	for the representations of auxiliary and actual momenta respectively, where ${\alpha}>0$ and ${\alpha}\neq1$.  
	
	Note that, as the wavefunctions $\psi(x)$ and $\phi(q)$ are connected through Fourier transformation,  they necessarily satisfy the Maassen--Uffink uncertainty relation~\cite{Maa88} 
	\begin{equation} \label{UP} %%%
	R_{\alpha}[\psi]+R_{\alpha^*}[\phi]\ge \ln\left(\pi\alpha^{\frac{1}{2(\alpha-1)}}{\alpha^*}^{\frac{1}{2(\alpha^*-1)}}\right)
	\end{equation}
	for conjugated indices, with $1/\alpha + 1/\alpha^*=2$. 
	Although this relation has been improved by considering the probability density governing the measurement process \cite{Ras17}, to the best of our knowledge the correction to this inequality taking into account the transformation~\eqref{transformation} has not been developed until now, except for the Shannon case~\cite{Ras17} that corresponds to the limit $\alpha=\alpha^*=1$. 
	Notice that,  from Eqs.~\eqref{transformation} and~\eqref{Renyigamma}, it follows trivially that $R_{\alpha}[\tilde\phi_n]>R_{\alpha}[\phi_n]$ whenever $\phi$ is not the Dirac's delta. 
	
	As an example we show in Table~\ref{tableRenMom} the behavior of the R\'enyi entropies corresponding to the ground state and first five excited ones in both representations of momentum space, for $\alpha=2$, and for different values of the deformation parameter~$\beta$. 
	\begin{table}[H]
		\caption{Numerical computation of R\'enyi entropy $R_2$, in auxiliary (left) and actual (right) momentum space, for the ground and first 5 excited  states of the 1D harmonic system with minimal length.}
		\label{tableRenMom}
		\centering
		\begin{tabular}{cccc|ccccc}
			\hline
			\textbf{$R_2[\phi_{n}]$} & \textbf{$\beta=0.1$} & \textbf{$\beta=0.5$} &   \textbf{$\beta=1$} \hspace{8mm} & \quad \  \quad &  \textbf{$R_2[\tilde\phi_{n}]$} & \textbf{$\beta=0.1$} & \textbf{$\beta=0.5$} & \textbf{$\beta=1$} \\
			\hline
			\midrule
			$n=0$& 0.876 & 0.723 & 0.565 \hspace{8mm} && $n=0$& 0.899& 0.808 & 0.690  
			\\
			$n=1$ & 1.119 	& 0.859  & 0.640\hspace{8mm} 	&&$n=1$ & 1.229 	& 1.227  & 1.153	 
			\\
			$n=2$ & 1.242 	&  0.916  &  0.669\hspace{8mm} && 	$n=2$ & 1.432 	&  1.424  &  1.285 
			\\
			$n=3$ & 1.322 & 0.949  & 0.685\hspace{8mm} &&			$n=3$ & 1.582 & 1.533  & 1.341
			\\
			$n=4$ & 1.380 & 0.971 & 0.695\hspace{8mm} && 			$n=4$ & 1.700 & 1.599 & 1.370
			\\
			$n=5$ & 1.424 & 0.987 &  0.701\hspace{8mm} && 			$n=5$ & 1.798 & 1.642 &  1.388
			\\
			\hline
			\bottomrule
		\end{tabular}
	\end{table}
	
	In Table~\ref{tableRenPos} 
	we show some particular values of the position-momentum R\'enyi entropies' sum for different values of the parameter $\beta$, various quantum states of the harmonic system with minimal length, and fixed $\alpha=\frac23$, then $\alpha^*=2$. 
	Note that, as expected,  all values are bigger than the lower bound in~\eqref{UP},  given by \ $\ln\left(\frac{3\sqrt 3\,\pi}{2}\right) \simeq 2.100$ %2.0995$ 
	for these particular values of the entropic parameters $\alpha$ and $\alpha^*$. 
	\begin{table}[H]
		\caption{Numerical computation of R\'enyi entropy %$R_2[\psi_{n}]$ 
			in position space (left) and sum of position--momentum entropies (right), for the ground and  first 5 excited states of the 1D harmonic system with minimal length.}
		\centering
		%% \tablesize{} %% You can specify the fontsize here, e.g.,  \tablesize{\footnotesize}. If commented out \small will be used.
		\begin{tabular}{cccc|ccccc}
			\hline
			\textbf{$R_2[\psi_{n}]$} & \textbf{$\beta=0.1$} & \textbf{$\beta=0.5$} & \textbf{$\beta=1$}\hspace{6mm} & \qquad &  \textbf{$R_\frac23[\psi_{n}]+R_{2}[\tilde\phi_{n}]$} & \textbf{$\beta=0.1$} & \textbf{$\beta=0.5$} & \textbf{$\beta=1$} \\
			\hline
			$n=0$& 0.974 & 1.167  & 1.356 \hspace{6mm} &&  $n=0$ & 2.123 & 2.205 & 2.290
			\\
			$n=1$ & 1.276 	&  1.506  & 1.717 \hspace{6mm} &&   $n=1$ &  2.723	&  2.939&  3.106
			\\
			$n=2$ &1.426	& 1.623& 1.819\hspace{6mm} &&  $n=2$ & 3.084	&  3.307  &  3.417
			\\
			$n=3$ & 1.517 & 1.668  & 1.855\hspace{6mm} && $n=3$ & 3.346 & 3.526  & 3.585
			\\                                           
			$n=4$ & 1.576 &1.684 & 1.868 \hspace{6mm} && $n=4$ & 3.552 & 3.673 &  3.693
			\\			
			$n=5$ & 1.615 & 1.688 & 1.872 \hspace{6mm} && $n=5$ & 3.719 & 3.776 &   3.770
			\\
			\hline
		\end{tabular}
		\label{tableRenPos}
	\end{table}

	%%%%%%%%%%%%%%%%%%%%%%%%%%%%%%%%%%%%%%%%%%
	\section{Discussion}
	
	In these proceedings we show a numerical analysis of informational measures such as R\'enyi entropies and their sum, in the case of the 1-dimensional quantum harmonic oscillator wavefunctions assuming for the position and momentum operators a deformed commutation relation, characterized by a parameter $\beta$. A nonvanishing  deformation parameter implies the existence of a minimal length, which is proposed to be a characteristic of quantum gravity theory. Further findings for arbitrary pairs of entropic indices below the  conjugacy curve, could also be obtained, and will be presented elsewhere together with a comparison with known lower bounds for the entropies' sum. Future work includes consideration of other physical systems and/or a more general deformed commutator between position and momentum observables (in $D$ dimensions), focusing on those states that minimize the generalized uncertainty relations.

	%%%%%%%%%%%%%%%%%%%%%%%%%%%%%%%%%%%%%%%%%%
	\vspace{6pt} 
	
	%%%%%%%%%%%%%%%%%%%%%%%%%%%%%%%%%%%%%%%%%%
	\subsection*{authorcontributions}{Both authors have contributed to conceptualization, methodology, formal analysis, and writing--original draft preparation. }
	
	%%%%%%%%%%%%%%%%%%%%%%%%%%%%%%%%%%%%%%%%%%
	\subsection*{funding}{This research was funded by Universidad Nacional de La Plata (UNLP), Argentina, project number 11/X812. MP is grateful also to Consejo Nacional de Investigaciones Cient\'ificas y T\'ecnicas (CONICET), Argentina, for financial support. DPC acknowledges Universidad de Granada, Spain.}
	
	%%%%%%%%%%%%%%%%%%%%%%%%%%%%%%%%%%%%%%%%%%
	\subsection*{conflictsofinterest}{The authors declare no conflict of interest. The funders had no role in the design of the study; in the collection, analyses, or interpretation of data; in the writing of the manuscript, or in the decision to publish the results.}

	%=====================================
	% References, variant A: internal bibliography
	%=====================================

\end{document}